\documentclass[10pt,letterpaper]{article}
\usepackage[top=0.85in,footskip=0.75in]{geometry}

\usepackage{changepage}

\usepackage[utf8]{inputenc}

\usepackage{textcomp,marvosym}

\usepackage{fixltx2e}

\usepackage{amsmath,amssymb}

\usepackage{cite}

\usepackage{nameref,hyperref}

\usepackage{microtype}
\DisableLigatures[f]{encoding = *, family = * }

\usepackage{rotating}

\usepackage[aboveskip=1pt,labelfont=bf,labelsep=period,singlelinecheck=off]{caption}

\makeatletter
\renewcommand{\@biblabel}[1]{\quad#1.}
\makeatother

\date{}

\newcommand{\ave}[1]{\left \langle #1 \right \rangle}
\newcommand{\col}[1]{\left \{ #1 \right\}}

\begin{document}
\vspace*{0.35in}

\begin{flushleft}
{\Large
\textbf\newline{Supersampling and Network Reconstruction of Urban Mobility}
}
\newline
\\
Oleguer Sagarra\textsuperscript{1,2,*},
Michael Szell\textsuperscript{2,3},
Paolo Santi\textsuperscript{2,4},
Albert D\'iaz-Guilera\textsuperscript{1},
Carlo Ratti\textsuperscript{2},
\\
\bigskip
\bf{1} Departament de F\'{\i}sica Fonamental, Universitat de Barcelona, Barcelona, Spain
\\
\bf{2} Senseable City Laboratory, Massachusetts Institute of Technology, Cambridge, Massachusetts, USA
\\
\bf{3} Center for Complex Network Research, Northeastern University, Boston, Massachusetts, USA
\\
\bf{4} Istituto di Informatica e Telematica del CNR, Pisa, Italy
\bigskip


* osagarra@ub.edu

\end{flushleft}
\section*{Abstract}
Understanding human mobility is of vital importance for urban planning, epidemiology, and many other fields that draw policies from the activities of humans in space. Despite the recent availability of large-scale data sets of GPS traces or mobile phone records capturing human mobility, typically only a subsample of the population of interest is represented, giving a possibly incomplete picture of the entire system under study. Methods to reliably extract mobility information from such reduced data and to assess their sampling biases are lacking. To that end, we analyzed a data set of millions of taxi movements in New York City. We first show that, once they are appropriately transformed, mobility patterns are highly stable over long time scales. Based on this observation, we develop a {\em supersampling} methodology to reliably extrapolate mobility records from a reduced sample based on an entropy maximization procedure, and we propose a number of network-based metrics to assess the accuracy of the predicted vehicle flows. Our approach provides a well founded way to exploit temporal patterns to save effort in recording mobility data, and opens the possibility to scale up data from limited records when information on the full system is required.


\section*{Introduction}
The increased pervasiveness of information and communication technologies is enabling the tracking of human mobility at an unprecedented scale. Massive call detail records from mobile phone activities \cite{gonzalez2008uih,blondel2015} and the use of global positioning systems (GPS) in large vehicle fleets \cite{bazzani2010slu} for instance, are generating extraordinary quantities of positional and movement data available for researchers who aim to understand human activity in space. 
Other data sources, such as observations of banknote circulation \cite{brockmann2006slh, thiemann2010sbs}, online location-based social networks \cite{scellato2011ssp, scellato2011nps}, radio frequency identification traces \cite{barthelemy2010sn, cattuto2010dpi, roth2010cpc}, or even virtual movements of avatars in online games \cite{szell2012ums} have also been used as proxies for human movements. These studies have provided valuable insights into several aspects of human mobility, uncovering distinct features of human travel behavior such as scaling laws \cite{brockmann2006slh, song2010msp} or predictability of trajectories \cite{song2010lph} among others. Besides empirical studies, the surge of available data on human mobility has also evoked interest in developing new theoretical models of mobility at several scales. Such models have deep implications for various subjects ranging from epidemiology to urbanism \cite{hufnagel2004fce, belik2011nhm, colizza2006rat,balcan2009mmn}, with special importance in city planning and policy action \cite{batty2013nsc}. 

Despite these first success stories, the theoretical development of tools and techniques for handling massive data sets of human mobility and for assessing their possible biases is still a road full of obstacles. Existing models based on gravity \cite{zipf1946p}, radiation \cite{simini2012umm}, intervening opportunities \cite{stouffer1940iot}, etc.~present a first step towards an accurate proxy for mobility at medium and large range scales, but they have been proven to be not always satisfactory to describe short scale movement such as intra-city displacements. The size of the data analyzed, the multiple scales involved, the highly skewed statistical nature of human activities \cite{Clauset2009} and the lack of strict control on the reliability of the data 
are just some of the multiple challenges this exciting new era poses.

Although they are often extensive, one of the main limitations of data sets used in empirically driven urban-scale mobility research is the limited coverage of the entire population under study. 
For instance, cell phone data records are typically obtained from a single operator.
Similarly, data from taxis, or from other vehicle fleets 
are typically obtained from a single company, which usually represents only a small fraction of the actual number of vehicles circulating in a city \cite{bazzani2010slu,sagarra2011sca}. In some scenarios, fully grasping a certain mobility-related phenomenon 
may require modelling the entire population of interest. For instance, it was shown that the fraction of taxi trips that can be shared in the city of New York is an increasing (albeit not simple) function of the number of daily taxi trips \cite{Santi14}. Hence, if a certain data set covers only a fraction of the daily taxi trips performed in a city, the taxi sharing potential cannot be fully unveiled.

The above discussion motivates the need of extrapolating urban mobility data starting from a subset of the population of interest. Although a number of urban mobility studies have applied such methods\cite{Piorkowski09,Spieser14}, a definition and assessment of a statistically rigorous extrapolation methodology is so far lacking. Even the sub-problem of 
assessing the quality of urban movement models is to date open, since the skewness of the underlying statistical distributions \cite{roth2010cpc} makes a set of consistent, quantitative indicators hard to develop. 
In this paper, we fill these gaps by introducing a rigorous methodology to tackle the problem of obtaining an accurate picture of a mobility process when only a limited observation of such a process is available, both in time and volume.
We first propose a simple rescaling rule which allows to quantify the strong temporal regularity of urban mobility patterns, even at very fine scales such as trips between particular intersections.
Exploiting this regularity, we use a maximum entropy approach combining empirical data to model the occurrence of the core of frequent trips with an exponential gravity model \cite{wilson1970stp,erlander1990gmt,sagarra2013smm} accounting for the variation observed in the least-frequent trips.
We apply our method to accurately reconstruct the data set of all taxi trips performed in the city of New York in the year 2011 using small fractions sub-sampled from only a month of recorded data. By analysing the temporal patterns and the topological properties of the yearly mobility of taxis represented as a multi-edge network, we can finally assess the statistical accuracy of the proposed {\em supersampling} methodology using a number of both information-theoretical and network-based performance metrics.

The remainder of the paper is structured as follows: We first present the study of both the temporal and topological patterns observed in the data, which then allows us to construct a maximum entropy method that exploits these features to solve the {\em supersampling} problem. Finally, we systematically test our reconstruction model on a very large data set and conclude by discussing some insights about the structure of urban mobility that the present study draws.

\section*{Results}
Typically, mobility data is formalized by so-called Origin-Destination (OD) matrices, which are particular examples of weighted, or multi-edge networks \cite{sagarra2014efn}. OD matrices represent the number of observed trips $\hat{t}_{ij}$ between the $L=N^2$ pairs of $N$ locations or nodes $i,j$ over a given observation period $\tau$. A location can be defined based on a spatial partitioning of the urban area, on points of interest \cite{Peng2012chm}, or on road intersections \cite{Santi14} -- as it is the case in the NY taxi data set at hand (see methods). Given this network representation, one can compute the total incoming $\hat{s}^{in}_j = \sum_{i} \hat{t}_{ij}$ and outgoing strength $\hat{s}^{out}_i = \sum_{j} \hat{t}_{ij}$ of a node $i$.
Throughout this paper, we define {\em active nodes} as the subset of nodes which are either origin or destination of at least one trip in the set of all recorded trips $\hat{T} = \sum_{ij} \hat{t}_{ij}$
and similarly {\em active edges} as the pairs of locations with at least one trip ($\hat{t}_{ij}\geq1$) recorded between them. The notation $\hat{x}$ shall refer to the observed value of the random variable $x$ as derived from the data set, $\ave{x}$ to its expected value over independent realizations of a given model, while $\bar{x}$ denotes the matrix or network average of the variable $\hat{x}$ across the full empirical OD matrix (for example, average graph-degree $\bar{k}$). Finally, the symbol $\ave{x}_{\tau}$ is used to express averages over time of variable $x$ using bins of temporal length $\tau$.

\subsection*{Stability of temporal urban-mobility patterns}\label{sec:tempPatt}
\begin{figure}[tbp]
\centering
\includegraphics[width=\linewidth]{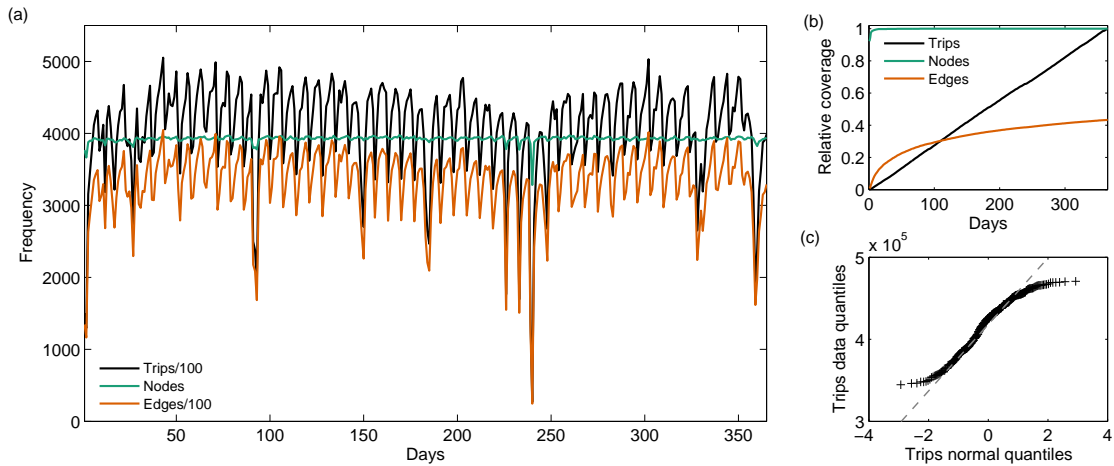}
\caption{\textbf{(a)} Circadian clocks in the city: Time-series of the number of observed trips $\hat{T}$, active edges $\hat{E}$ and active nodes in the city of NY per day over the year 2011. The influence of both seasonal fluctuations, major events and stable weekly patterns are clearly observed. \textbf{(b)} Aggregated fraction of active nodes and total trips as days of data are accumulated normalized by the total number of recorded nodes and trips at the end of the observation period, the evolution of the accumulated graph density in time $E/L = E/N^ 2$ (observed binary edges over total possible pairs of intersections) is also reported. Nodes are almost fully sampled within the first days analyzed, while edges are sampled sub-linearly in time. \textbf{(c)} Quantile - Quantile plot comparing the number of trips per day observed in the data set without outliers within two standard derivations of the mean ($>95\%$ of the data) to the theoretical quantiles of a normal distribution and linear fit (dashed line) showing their proportionality (similar results not shown obtained for number of edges and nodes respectively).
}
\label{fig1}
\end{figure}

While the built structure of cities evolves slowly in time, many dynamic, behavioral processes that take place within a city unfold relatively fast, and in principle could be strongly variable across time. However, human activity in cities exhibits highly regular patterns when observed over well defined periods of time, such as circadian or weekly rhythms. Intra-urban mobility is a good example for such activities: With longer time spans or larger samples of gathering movement data in cities, the picture of the underlying mobility network will clear up continually, but stable patterns should already emerge with relatively few data points as we can see in Fig.~\ref{fig1}A. To systematically test this hypothesis, we make use of a fleet of taxis acting as probes, sampling from the total traffic of all vehicles in a city. The total number of recorded trips, or sampling size $\hat{T}(\tau)$, depends on the total observation time $\tau$ and the number of probes,
i.e., the size of the sub-population that is being monitored. 

The evolution of the sampling size of trips as a function of the observation period, Fig.~\ref{fig1}B, can be extremely well approximated by a linear relation ($R^2>0.999$), indicating that the total number of trips generated daily in the city can be described as a random variable strongly concentrated around its mean value $\ave{T}_{\tau=1\,\mathrm{day}} \simeq 403,000 \pm  61,000$ (confidence bounds reported as standard deviation).

On a yearly scale, the distribution of trips per day $T_{\tau=1\,\mathrm{day}}$ is not statistically compatible to a Gaussian distribution mainly due to seasonal effects, see Fig.~\ref{fig1}A. The effects of summer and winter holidays are apparent, and of Hurricane Irene that hit New York city towards the end of August, but if we disregard such outliers, corresponding to around $5\%$ of the data that lies further than two standard deviations away from the mean,
the quantile-quantile plot shows acceptable agreement with a normal distribution, see Fig.~\ref{fig1}C.

To observe whether this strong regularity is also present in the finer structure of mobility, we must focus on each of the $N$ nodes and $L$ intersection pairs. Yet, we must find a suitable scaling to the data: The accumulated observed strength (both incoming and outgoing) $\hat{s}_i(\tau)$ of each node and the weight $\hat{t}_{ij}(\tau)$ of each intersection pair will increase as more and more data is gathered,
but if we normalize their (in or out) strength and weight by the total number of observed trips in the period $\tau$, a strong regularity is recovered as we show in the following. The quantities $\hat{p}_s^{out,in}$ and $\hat{p}_{ij}$, which quantify the {\em relative importance} of a given node and intersection pair compared to the overall network,
\begin{equation}\label{eq:rel_str}
\hat{p}_i^{out,in} (\tau) = \frac{\hat{s}^{out,in}_i (\tau)}{\hat{T}(\tau)} \qquad 
\hat{p}_{ij} (\tau) = \frac{\hat{t}_{ij} (\tau)}{\hat{T}(\tau)}
 \end{equation}
are extremely stable as shown in Table~\ref{table:node_stab}. We have split our data set into $n_{\tau}$ equal time intervals (on daily, weekly and four-week bases) and computed the relative dispersion of the values accumulated over the entire data set $\hat{p}(\tau=\tau_{max} = 1\,\mathrm{year})$ around the measured values $\ave{\hat{p}}_{\tau}$,
\begin{equation}\label{eq:rel_error}
\varepsilon = \frac{\hat{p}(\tau_{max})-\ave{\hat{p}(\tau)}_\tau}{\ave{\hat{p}(\tau)}_\tau},
\end{equation}
where $\tau_{max}$ is the time at the end of the full observation period and the averages are performed over all the time slices of length $\tau$. The graph-average of $\varepsilon$ is very close to zero and highly concentrated around this value for all the time windows considered (with a standard deviation of $13\%$ in the worst case, decaying as sampling time is increased). 

\begin{figure}[tbp]
\centering
\includegraphics[width=0.9\linewidth]{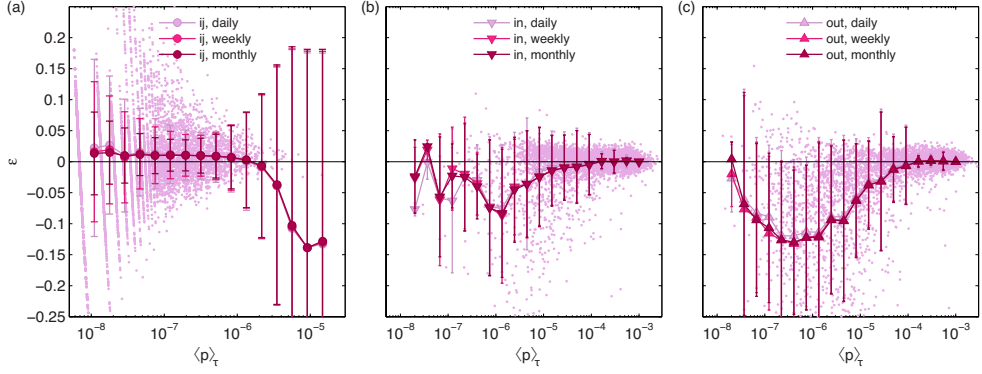}
\caption{The effect of sampling on node and intersection pair temporal stability. Correlation between measured values of intersection pair $\ave{\hat{p}_{ij}}$ (a) and node statistics $\ave{\hat{p}_s^{in}}$ (b), $\ave{\hat{p}_s^{out}}$ (c) for different time slices and relative dispersion around the mean (equation \eqref{eq:rel_error}) for the yearly aggregated data.
Error bars represent standard deviations on the log-binned data. Raw data for the daily case is shown in the background. For visual clarity, panel a) only shows a random subsample of $1/1000$ of the original points.}
\label{fig2}
\end{figure}

Fig.~\ref{fig2} shows the correlation between the relative error and the relative importance of nodes and links. The fact that $\sum_{ij} \hat{p}_{ij} = \sum_{i} \hat{p}_i^{out} = \sum_i\hat{p}_i^{in}=1$, coupled with second order seasonality effects 
induces an uneven distribution of errors: An overestimation of some values in the collection $\{\ave{p_{ij}}\}$ will forcefully induce an underestimation in some other values of the collection. Despite this issue, we can clearly see that the vast majority of the mass of relative errors is concentrated around zero (see points in background for Fig.~\ref{fig2}).

\begin{table}[h]
\resizebox{\linewidth}{!}{%
\begin{tabular}{c|| c c | c c | c c }
Time windows $n_\tau$&$\hat{p}_s^{in}$:  $\bar{\varepsilon}_{in} $ ($\pm$ std) &Outliers & $\hat{p}_s^{out}$: $\bar{\varepsilon}_{out} $ ($\pm$ std) &Outliers& $\hat{p}_{ij}$: $\bar{\varepsilon}_{inter} $ ($\pm$ std) &Outliers \\\hline
347 (1 day period)&
$-0.009\pm0.056$&$0.039$&$-0.037\pm0.095$&$0.083$&$0.018\pm0.127$&$0.025$\\
51 (1 week period)&
$-0.011\pm0.054$&$0.041$&$-0.040\pm0.095$&$0.088$&$0.014\pm0.100$&$0.026$\\
13 (4 weeks period)&
$-0.011\pm0.054$&$0.041$&$-0.040\pm0.095$&$0.087$&$0.012\pm0.061$&$0.025$\\
\end{tabular}
}
\caption{Variability of node and node-pair statistics (incoming $\hat{p}_s^{in}$ and outgoing $\hat{p}_s^{out}$ relative strength, and relative number of trips between intersections $\hat{p}_{ij}$) using different temporal granularity of the data set averaged over the full network (nodes and pairs of nodes respectively) compared to final yearly values. Time units with a total number of trips at least two standard deviations apart from the adjusted yearly mean have not been considered in the average to account for seasonal variations ($<5\%$ of the data in the worst case). For the pairs of intersections $ij$, only pairs with at least one non-zero appearance on the time slicing have been considered for the average. The fraction of data with absolute relative error larger than two standard deviations is also reported as \textit{Outliers}.
}
\label{table:node_stab}
\end{table}

To some extent, we would expect the node strength to be stable over time, since the number of trips received and generated at each location depends on parameters such as population density, number of points of interest present in a given location, etc.~\cite{yang2014lpc}, whose evolution is given by much slower dynamics than the mobility process studied herein. But additionally, time stability is also observed at the trip level between intersections, yet in this case   
the analysis displays a higher variability around the mean -- see Fig.~\ref{fig2}A and Table~\ref{table:node_stab}. This higher variability can be explained by the different sampling processes: While the percentage of active nodes becomes extremely stable already when just a very small number of days is considered (Fig.~\ref{fig1}B), this is not the case for the total number of active edges, because sampling of edge-specific attributes requires to grow as $L\simeq N^2$ to achieve a comparable level of accuracy. 

\subsection*{The anatomy of urban flows}
\begin{figure}[tbp]
\includegraphics[width=\linewidth]{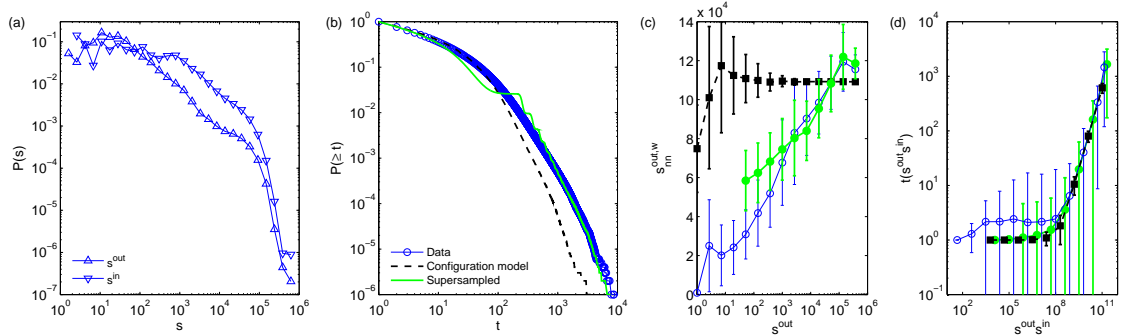}
\caption{Empirical network features of the taxi multi-edge mobility network. \textbf{(a)} Distribution of incoming and outgoing strengths $s$. \textbf{(b)} Existing edge weight complementary cumulative distribution function compared with a configuration model \cite{sagarra2014efn} and the supersampled model with $f=0.1$. \textbf{(c)} Weighted average neighbor out-strength $s^w_{nn}$ as function of node out-strength (similar results for incoming strengths not shown). \textbf{(d)} Graph-average existing weight as a function of product of outgoing and incoming strengths of origin and destination for a single instance of the network. Points represent mean values and error bars represent standard deviations computed using log-binning and distributions also shown using log-binning and ensemble results averaged over 100 repetitions of the two models.}
\label{fig3}
\end{figure}

The results on temporal stability indicate that any model aiming to reproduce human mobility at urban scales should consistently exhibit regularities as reported above. Having seen that the mobility patterns are stable across time, an understanding of the main topological aspects of the aggregated static picture is further needed in order to be able to select the main features our methodology should aim to reproduce.
The most relevant topological aspect of the mobility network is the highly skewed concentration of taxi pick-ups and drop-offs across the city, which gives rise to heavy-tailed node-strength distributions, Fig.~\ref{fig3}A (other general metrics are reported in the methods section).
Therefore, to test whether this relevant property alone already captures the essential features of the mobility network, we must consider a null model which randomizes the considered network keeping constant the strength of each node --the  multi-edge configuration model (see methods section or \cite{sagarra2014efn} for more details).
When comparing the null model with the data, we observe significant deviations showing the importance of certain places or nodes in the network. In other words, even the strong heterogeneity of the distribution of strengths cannot account for the skewness of the weight distribution: Additional factors add many more trips between some connections than there should be under random conditions. The empirical link weight distribution, blue line in Fig.~\ref{fig3}B, is more skewed than under a random allocation of trips (Configuration model), dashed line, and the connections at the node level show a clear assortative correlation instead of a flat profile, see Fig.~\ref{fig3}D. This occurs despite the fact that the average number of trips between the most busy locations can be characterized by the relation $\bar{t}\propto s^{out}_i s^{in}_j$ with a reasonable accuracy, see Fig.~\ref{fig3}C, coinciding with the configuration model. These insights indicate that
the distribution of node strengths across the city has a strong influence on the topology of the network, and needs to be taken into account when modelled, but needs extra ingredients to fully account for the observed pattern of connections. 

\subsection*{A flexible model to reproduce human mobility}
A general maximum entropy based theory for model generation (see \cite{sagarra2013smm,wilson1970stp,bianconi2009a} for extended discussion and references) allows us to efficiently exploit both the observed temporal stability features and the heterogeneous topological properties of the network to solve the {\em supersampling} problem at hand. It starts with the assumption that each intersection pair in the network is allocated a constant fraction $p_{ij}$ of the total sampling (equation \eqref{eq:rel_str}). Under this condition, and assuming that the mobility process is driven by some general constraints, such as population density or budget, it can be proved that for any desired level of sampling $T_{d}$ the statistics of trips for each pair of nodes can be well described by a set of $L$ independent Poisson processes with mean $\ave{t_{ij}} = T_d \ave{p_{ij}}$, with
\begin{equation}
\ave{p_{ij}} \equiv \lim_{\tau \to \infty}  \ave{\frac{t_{ij}(\tau)}{T(\tau)}} =  \lim_{T \to \infty}  \ave{\frac{t_{ij}}{T}} \qquad T \equiv \sum_{ij} t_{ij}.
\end{equation}

Following the theory, it would seem clear that from knowing the \textit{real} values of the collection  $\col{\ave{p_{ij}}}$, {\em supersampling} a mobility data set would be a trivial operation of generating $L$ independent Poisson processes using the provided proportionality rule. Therefore, the problem now reduces to inferring the collection of values $\{p_{ij}\}$ from an available data set. We shall assume that only one snapshot of the aggregated mobility network is available to this end (thus assuming no temporal information is available on the trip data) as is usually the case in mobility studies. The maximum likelihood estimation of such values corresponds to
\begin{equation}\label{eq:max_L_p}
\ave{\hat{p}_{ij}}_{ML} =  \frac{\hat{t}_{ij}(\tau)}{\hat{T}(\tau)} = \hat{p}_{ij}(\tau).
\end{equation}
There is, however, a practical issue in this formula related with the normalization condition for the random variables $\{p_{ij}\}$ and the presence of empty intersection pairs in the available observed data. For such intersections, using the formulas above, we have that $\hat{p}_{ij} = \hat{t}_{ij} = 0$, whereas their {\em real} $\ave{p_{ij}}$ value is unknown but fulfils $\ave{p_{ij}} \in [0,\hat{p}_{min} \sim \hat{T}^{-1}]$. Since by definition both collections $\{p_{ij}\}$ and $\{\hat{p}_{ij}\}$ need to be normalized, and denoting the set of active edges as $\mathcal{E} = \col{ij | \hat{t}_{ij}>0}$, we have,
\begin{equation}\label{eq:norm}
\sum_{ij} \ave{p_{ij}} = 1 = \sum_{ij | ij \in \mathcal{E}}  \ave{p_{ij}} + \sum_{ij | ij \notin \mathcal{E}}  \ave{p_{ij}} \qquad\qquad \sum_{ij} \hat{p}_{ij} = \sum_{ij| ij \in \mathcal{E}} \hat{p}_{ij} + 0 = 1,
\end{equation}
from which we see that $\sum_{ij |\in \mathcal{E}} \ave{p_{ij}} \leq \sum_{ij |\in \mathcal{E}} \hat{p}_{ij}  = 1$.

Hence, in general we cannot consider the empirically observed probabilities $\hat{p}_{ij}$ as a good proxy for the \textit{real} values of $p_{ij}$ unless the number of empty intersection pairs is very reduced. Given that the percentage of active edges (pairs of nodes for which $\hat{t}_{ij}>0$) is a very slowly increasing function of the sampling, see Fig.~\ref{fig1}B, inferring directly the set of probabilities $\col{p_{ij}}$ empirically would take an enormous data set -- note that even with over a year of data only roughly $40\%$ of edges are covered. 

For the reasons given above, a simple proportionality rule using equation \eqref{eq:max_L_p} is not a good {\em supersampling} strategy, specially for skewed and sparse data sets.

\subsection*{Supersampling urban trips}\label{sec:model}
Based on the previous discussion, we now present the methodology for {\em supersampling} an urban mobility data set that consists in inferring the collection of $\{\ave{p_{ij}}\}$ values from a set of aggregated empirical trips $\col{\hat{t}_{ij}}$. We should do so bearing in mind that despite the maximum likelihood formula in equation \eqref{eq:max_L_p} cannot be directly used for the empty intersection pairs in the data, it does perform well for non-empty intersections (see Fig.~\ref{fig2}). 

The maximum entropy based framework naturally allows such a procedure, since it can combine any constraint driven model
with the rich information encoded in the trip sample.
We propose a method to predict trips based on the theory mentioned earlier: Taking the $L$ intersection pairs (being them active edges in the data set or not), we split them into two parts, the subgroup of \textit{trusted} trips defined as $\mathcal{Q} = \col{ij | \hat{t}_{ij} > t_{\min}}$ and its complementary part $\mathcal{Q}^C$. The value $t_{\min}$ is a threshold modelling a minimal statistical accuracy that depends on the amount of data available, and which may be set to 1 in practical applications.
We keep the proportionality rule $\hat{p}_{ij} \propto \hat{t}_{ij}$ for the {\em trusted} trips, while for the remaining trips we apply a doubly constrained exponential gravity model --other maximum entropy models \cite{sagarra2013smm} could also perform well as long as they preserve the outgoing and incoming strength.
In other words, we generate a collection of $\col{\ave{p_{ij}}}$ values,
\begin{equation}\label{eq:p}
\ave{p_{ij}} = \left \{ \begin{array}{l l}
\frac{\hat{t}_{ij}}{\hat{T}} = \hat{p}_{ij} & ij \in \mathcal{Q} \\
x_i y_j e^{-\gamma c_{ij}} & ij \in \mathcal{Q}^C 
\end{array} \right. .
\end{equation}
The values $\gamma$ and $\col{x_i,y_j}$ are the $2N+1$ Lagrange multipliers satisfying the following equations
\begin{eqnarray}\label{eq_saddle}
\hat{s}_i^{out} - \hat{T} \sum_{i|ij \in \mathcal{Q}} \hat{p}_{ij} & = &\hat{T} x_i \sum_{i | ij \in \mathcal{Q}^C} y_j e^{-\gamma c_{ij}} \nonumber \\
\hat{s}_j^{in} - \hat{T} \sum_{j|ij \in \mathcal{Q}} \hat{p}_{ij} & = & \hat{T}y_j \sum_{j | ij \in \mathcal{Q}^C} x_i e^{-\gamma c_{ij}} \nonumber \\
 \hat{C} -  \hat{T} \sum_{ij|ij \in \mathcal{Q}}   c_{ij} \hat{p}_{ij} & = & \hat{T} \sum_{ij|ij \in \mathcal{Q}^C}   c_{ij} x_i y_j e^{-\gamma c_{ij}} ,
\end{eqnarray}
where $\hat{C}=\sum_{ij} \hat{t}_{ij} c_{ij}$ is the total euclidean distance of the observed trips ($c_{ij}$ stands for the distance between intersections $i$ and $j$). Note that, by construction, the values are properly normalized, i.e., $\sum_{ij} \ave{p_{ij}} = \sum_{ij \in  \mathcal{Q}} \hat{p}_{ij} + \sum_{ij \in \mathcal{Q}^C} x_i y_j e^{-\gamma c_{ij}} = 1$. 

The model presented earlier needs to deal with the issue of inactive nodes that do not appear in the original data due to poor sampling, i.e., nodes for which $\hat{s}=0$ either incoming or outgoing for some observation period $\tau$. This issue has a minor impact in our case due to the previously observed rapid coverage of the number of active nodes 
(the number of inactive nodes is negligible after accumulating very few days of data, see Fig.~\ref{fig1}B). In any case, it can be solved easily: given that the geographic positions of the nodes are available, we could always artificially assign a certain relative strength to the nodes not present in the data using complementary call detail records\cite{Lenormand2014}, census data or points of interests (POI) data, or assign them some values according to a chosen distribution depending on the data at hand.
For simplicity, in our case we have chosen to keep only the nodes present in the original data.

\subsection*{Assessing the quality of the supersampling methodology}
To test the {\em supersampling} methodology, we have proceeded to select a timespan of our data set corresponding to an observation period of $\tau = 1$ month (February 2011) from which we further randomly sub-sample different fractions $f$ used as training sets to compute $\col{\ave{p_{ij}}}$ applying equations \eqref{eq:p} and \eqref{eq_saddle}. We then reconstruct the OD using the proportionality rule $\col{\ave{t_{ij}} = T_d \ave{p_{ij}}}$ for both the complete and reduced data set, $T_d = T(\tau' = \text{1 year})$ and $T_d=T(\tau = \text{1 month})$. Finally, we compare the model predictions with the set of empirically observed trips in these periods.

In order to do so, we need to introduce metrics to quantify the resemblance between model predictions and actual recorded values. Commonly used indicators are the Sorensen-Dice common part of commuters (CPC) value \cite{Lenormand2012} or the linear fit of $\ave{t_{ij}}^{model}$ vs $\hat{t}_{ij}$ \cite{simini2012umm}. However, the skewness of the observed trip distribution represents a challenge to these indicators: The variability of low-valued trips induces notable instabilities on both and being single numbers, they are only able to provide a limited picture on the precision of a given model to reproduce empirical results. 

To overcome these issues, we propose a slight modification to both the Sorensen CPC index and the coefficient of determination $R^2$ from the fit $\ave{t_{ij}}^{model}$ vs $\hat{t}_{ij}$ (see methods) and we additionally propose the introduction of a number of network metrics to precisely assess the quality of the models used in human mobility at the topological, finer, scale: The unweighted degrees of the nodes $\ave{k}(\hat{s})$, the weighted neighbor strength correlation $\ave{s^{w}_{nn}}(\hat{s})$, existing trip distribution $P(t)$ and number of existing trips as a function of the origin destination strength product $\bar{t}(s s')$.

\begin{table}[htbp]
\centering
\begin{tabular}{ c || c c  c c || c c  c }
&\multicolumn{4}{c||}{$\tau=1$ month}&\multicolumn{3}{c}{$\tau=1$ year}\\\hline\hline
 $f$ &$f_{\mathcal{Q}}$& $\mathcal{L}/\mathcal{L}_{Emp}$& $CPC$ & $R^2_{cond}$&$f_{\mathcal{Q}}$ & $CPC$ &$R^2_{cond}$\\ \hline \hline
1.00 		 		  & 0.8855 & 1.46 & 0.92 & 1.00 & 0.07090 & 0.78 & 0.91\\
0.75				      & 0.6417 & 1.64 & 0.83 & 0.98 & 0.05138 & 0.76 & 0.89\\
0.50 		 		  & 0.4014 & 1.88 & 0.77 & 0.94 & 0.03214 & 0.74 & 0.86\\
0.25 		 		  & 0.1711 & 2.26 & 0.68 & 0.83 & 0.01370 & 0.69 & 0.78\\
0.10 		 		  & 0.0492 & 2.64 & 0.60 & 0.65 & 0.00394 & 0.65 & 0.63\\
0.01 		 		  & 0.0012 & -    & 0.58 & 0.21 & 0.00010 & 0.66 & 0.26\\
0.005		          & 0.0003 & -    & 0.59 & 0.12 & 0.00002 & 0.66 & 0.18\\\hline
Configuration          & -     & -    & 0.57 & -0.87 &   -      & 0.64 & -0.22\\
Empirical              &  1    & 1    & 1    & 1 & 1        & 1    & 1 
\end{tabular}
\caption{Parameters for the validation of the methodology. See details on each indicator in the main text and in the methods section. The number of {\em trusted} trips fed to the model relative to the entire number of generated trips $f_{\mathcal{Q}} = \sum_{ij | ij\in \mathcal{Q}} \hat{t}_{ij} / T_d(\tau)$ is reported in column 3. The {\em Supersampled} models with different fractions $f$ are only generated using subsamples of the training set (1 month observation period). {\em Empirical} stands for the model generated using the empirical probabilities $\hat{p}_{ij}$ (equation \eqref{eq:rel_str}) of the full data set 
and {\em Configuration} stands for the multi-edge configuration model applied to the full data set.
}
\label{table:res_super}

\end{table}

\begin{figure}[tbp]
\centering
\includegraphics[width=\linewidth]{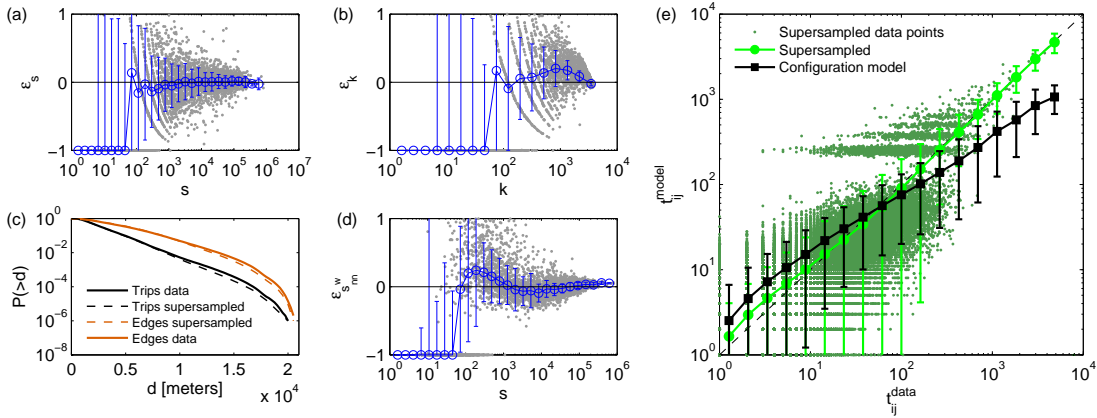} 
\caption{Main network differences between real data accumulated over a year and Supersampled model from real one month data with $f=0.1$ subsampling. \textbf{(a)-(d)} Relative error ($(\ave{x}-\hat{x})/\hat{x}$ with $\hat{x}$ being a magnitude measured from the aggregated yearly network) between reconstructed network using supersampling and original data for outgoing degrees (a), strengths (b) and average neighbor strength (d) (similar results for incoming direction not displayed). The complementary cumulative distribution function of both edge lengths and trips lengths (c) is also shown. \textbf{(e)} Comparison between empirical $\{\hat{t}_{ij}\}$ values and model prediction over a single run. Configuration model expectation from a single run using the full year data set is also shown for comparison. All results averaged over 100 repetitions of the model, error bars represent standard deviations on the log-binned data and raw data is shown in the background.
For visual clarity, panel e) only shows a random subsample of $1\%$ of the raw data in background.}
\label{fig4}
\end{figure}

The results for the supersampling method are summarized in Table~\ref{table:res_super} and a specific example for $f=0.1$ (reconstruction using only $10\%$ of the original data of the monthly data set compared to yearly data) is shown in Fig.~\ref{fig4} for the different indicators proposed. For comparison, results using both a configuration model
and the empirical values $\col{\hat{p}_{ij}}$ using the information encoded in the complete dataset are also shown.

We observe an accurate reconstruction of the mobility network for a wide range of values of $f$, which shows the validity of our proposed supersampling methodology. At the global scale, even at extreme levels of subsampling, our model is successful at reconstructing the original dataset. Also at the topological scale, despite the heterogeneities in the underlying distributions, the methodology generates very accurate predictions. The predictions for the least frequently visited nodes display higher relative errors due to the presence of inactive nodes in the training dataset ($1.6\%$ of total nodes for $f=0.1$). 

Upon close inspection, our inferred values $\col{\ave{p_{ij}}}$ slightly over-estimate low-valued weights and underestimate large-valued weights and strengths as was expected from the analysis in temporal stability (see Fig.~\ref{fig2}), yet the errors are greatly mitigated as we can see in Fig.~\ref{fig3}B. See Figs.~\ref{fig3}B (green line) and \ref{fig4}E (green dots) where we can observe a gap around $t \sim 100$ and $\ave{t_{ij}^+} \sim 100$, respectively, which corresponds to the separation between the trusted empirical data (separated points in the background belonging to the group of trusted trips $\mathcal{Q}$) and the reconstructed trips (clustered cloud of points). The minor seasonal fluctuations detected first in our temporal analysis together with these over- and under-estimations explain the minor limitations of the model to reproduce perfectly the entire yearly data set.

The second order effects induced by the seasonality of recorded data can also be seen in the performance of our methodology under extreme levels of subsampling (using around $1\%$ of the sample monthly data to feed the model or less). In these circumstances, the model is still able to produce a good prediction of the empirical data, yet it reproduces better the accumulated yearly mobility rather than the monthly one since the inherent seasonal variations of traffic between certain intersection pairs are smoothed by the aggregation procedure.

Furthermore, in the event that enough historical data were available, we could achieve even better results by computing the collection $\{\ave{p_{ij}}_\tau\}$ with an appropriate $\tau$ period (depending on the granularity of the data) and approximating $\ave{p_{ij}} \simeq \ave{p_{ij}}_\tau$ for the group of {\em trusted} trips (equation \eqref{eq_saddle}). Such a procedure, which may be extended to overcome the minor limitations imposed by the seasonality of the data and other improvements related with the presence of non-active nodes could be derived to perfect the method.

\section*{Discussion}
The stationarity of the temporal statistics of trips between different locations, 
together with a suitable scaling for the data observed over different timespans, has allowed us to develop a general model that is highly effective at reconstructing a general mobility scenario from very limited aggregated data, without the need of having fine grain temporal information, and provides insights about the structure of real taxi trips. The success of our reconstruction method, even using very small amounts of data, points out the composite structure of the network of urban mobility: Taxi displacements are characterized by a small core of very frequent trips coupled with trips generated at random but conditioned by the structural constraints of the city such as population distribution and mobility costs. 

Ultimately, the results presented in this paper could be used to answer questions that are of fundamental importance in the field of human mobility modelling, such as: $i)$ Can an accurate picture of urban mobility patterns be obtained from an incomplete sample of the population?, and $ii)$ are existing metrics sufficient to assess the quality of model predictions?

We pointed out the importance of data sampling and of the correct assessment of mobility models, and introduced network-based tools to evaluate such models. The implications of these findings are two-fold: On the one hand, the stationarity of the temporal patterns could be exploited to save space and effort in recording mobility data. On the other hand, our method opens the possibility of efficiently scale up data from reduced fleet of vehicles in cases where a full knowledge of the system is needed.

Our study provides a first step in showing that incomplete samples can indeed be scaled up adequately with the appropriate models, and that network metrics are required to comprehensively assess mobility model predictions.

\section*{Materials and Methods}
\subsection*{Data set}
	OD matrices are typically inferred from either census/survey data or alternative means such as social media data or Call Detail Records (CDR) \cite{robillard1975eom}. They must be constructed in terms of {\em trips}, i.e.~well defined trajectories between a starting and ending point. For this reason, we use the taxi data for its completeness, since all trips are recorded, and we consider it a good proxy for general urban mobility, with the caveat that commuting patterns and areas with little taxi demand are covered less. For the present paper we have used the full set of taxi trips (with customers) with starting and ending points within Manhattan obtained from the New York Taxi and Limousine Commission for the year 2011 via Freedom of Information Law request \cite{Santi14}. We have aggregated the data at the node level taking into account the grid of roads up to secondary level and adding trips to the nearest node (intersection) in a radius of $200 m$. We have decided to keep the self-loops present in the data for simplicity (albeit their fraction is completely negligible). In the analysis, all trips including both week-ends and week-days are considered, since the pattern for weekly trips shows a continuous increase in the number of trips peaking on Friday and followed by a sudden drop on Sundays.
 
For the subsampling of the reduced data set used as basis to reconstruct the networks, we have used random subsampling. The parameters of the obtained mobility networks for two different observation periods $\tau$ are reported in Table~\ref{table:net_feat}.

The analyzed data shows that most of the taxis share similar performance. This, coupled with the fact \cite{Piorkowski09} that individual taxi mobility traces are in large part statistically indistinguishable from the overall population, justifies that their individual traces (corresponding to sets of trips performed by different customers which can be considered as independent events) can be safely aggregated for the analysis.

\begin{table}[htbp]
\centering
\resizebox{\linewidth}{!}{%
\begin{tabular}{c | c c c| c c c |c  c c |c c | c c|c}
$\tau $& N&$\hat{E}$&$\hat{T}$&$\bar{s}$&$\sigma_{s}^{out}$&$\sigma_{s}^{in}$&$\bar{k}$&$\sigma_{k}^{in}$&$\sigma_{k}^{out}$&$\hat{T}/L$&$\hat{E}/L$&$\bar{t}$&$\sigma_{t}$&$\bar{c} = \sum d_{ij} \hat{t}_{ij}/\hat{T} $\\\hline
February 2011 & 4085&3099271&11768911& 2881& 4423&4112 &759&730&597&0.7&0.19&3.8&6.9&$2434\pm1801$m \\
Year  & 4091&7251605&146986835&35929&51325&54877&1773&799&1206&8.8&0.43&60.84&20.27&$2458\pm1831$m
\end{tabular}
}
\caption{Empirical network parameters. Symbol $\bar{x}$ stands for graph averaged magnitudes. See main text and methods for a definition of each magnitude.}
\label{table:net_feat}
\end{table}

\subsection*{Null model: The Multi-Edge Configuration model}
Cities usually display a high level of variability across its different locations in terms of activity, i.e., city center concentrate busy locations while outskirts usually display less traffic/retails areas and others. At the level of networks, this translates in important topological heterogeneities which need to be accounted for using a suitable null model. The Multi-Edge configuration model \cite{sagarra2014efn} is a maximum entropy model that is used to generate maximally random network instances of graphs with a prescribed strength sequence $\{s^{in},s^{out}\}$ (number of trips \emph{entering} and \emph{leaving} each node). The expected number of trips between two nodes $i$ and $j$ with respective strengths $s_i^{out}$ and $s_j^{in}$ reads $\ave{t_{ij}^{Conf}} = \hat{s}^{out}_i \hat{s}_j^{in}/\hat{T}$ and the assortativity profile is flat (nodes are uncorrelated at the level of strengths). Throughout this paper, we use this null model as benchmark preserving the strength sequence obtained from aggregating the complete yearly observation period.

	\subsection*{Indicators for the quality of the reconstruction}
		\paragraph{Distance based measures}
			\begin{itemize}
			\item \textbf{Sorensen-Dice common part of commuters index:} This indicator was proposed in \cite{Lenormand2012} and based on the original formulation is defined as
			\begin{equation}\label{eq:cpc_sample}
				CPC_{sample} = \frac{2 \sum_{ij \in \mathcal{E}} \min(\hat{t}_{ij},t^{model}_{ij})}{\sum_{ij} \hat{t}_{ij} + \sum_{ij\in\mathcal{E}} t_{ij}^{model}}.
			\end{equation}
			We propose an alternative version formulated in terms of averages which reads,
			\begin{equation}
			CPC = \frac{2 \sum_{ij \in \mathcal{E}} \min(\hat{t}_{ij},\ave{t_{ij}})}{\sum_{ij} \hat{t}_{ij} + \sum_{ij\in\mathcal{E}} \ave{t_{ij}}}.
			\end{equation}			
The different versions of this indicator have values in the range $[0,1]$, where $CPC=1$ indicates total coincidence between data and model and $CPC=0$ total disagreement. However, for sparse data sets with a skewed distribution of $\col{\hat{t}_{ij}}$ values, equation \eqref{eq:cpc_sample} may return values excessively lower than 1, even for models very close to reality. To exemplify this fact, Table~\ref{table:cpcs} shows a comparison of the performance of the two indicators for the models presented in Table~\ref{table:res_super}. For the {\em Empirical} model, we can see that the second version of the indicator recovers values close to 1 as would be expected. Furthermore, both indicators converge to very similar values as sampling is increased (the yearly data set contains roughly $12$ times more trips than the monthly one). 

\begin{table}[htbp]
\centering
\begin{tabular}{ c  | c c | c c }
& \multicolumn{2}{c|}{$\tau=1$ month}&\multicolumn{2}{c}{$\tau=1$ year}\\\hline
$f$ &$\ave{CPC_{sample}}$ & $CPC$& $\ave{CPC_{sample}}$ & $CPC$ \\ \hline
1.00 & $0.786$   & 0.92 & $0.777$ & 0.78\\
0.75 & $0.757$   & 0.83 & $0.758$ & 0.76\\
0.50 & $0.713$   & 0.77 & $0.731$ & 0.74\\
0.25 & $0.639$   & 0.68 & $0.686$ & 0.69\\
0.10 & $0.562$   & 0.60 & $0.647$ & 0.65\\
0.01 & $0.539$   & 0.58 & $0.655$ & 0.66\\
0.005& $0.543$   & 0.59 & $0.655$ & 0.66\\\hline
Configuration          & $0.525$ & 0.57 & $0.641$& 0.64\\
Empirical              & $0.829$ & 1 & $0.936$ & 1 \\
\end{tabular}
\caption{Values for the different versions of the common part of commuters index for the reconstructed models. $\ave{CPC_{sample}}$ and averaged trip values computed over 1000 repetition of each model, standard deviations lower than $10^{-3}$ for all cases. Note how differences between indicators disappear with increased sampling.
}
\label{table:cpcs}
\end{table}
			\item \textbf{Linear correlation $\ave{t_{ij}}^{model}$ vs $\hat{t}_{ij}$}: This method is widely used \cite{simini2012umm}. We report the coefficient of determination $R^2_{cond}$ in all tables, based on the comparison between real data and conditional values of the model on the existing edges (since we are using a biased statistic based only on the observed trips in the original data, not the entire set of intersection pairs). With Poisson distributed variables with mean $\ave{t_{ij}}$, such conditioned average reads,
\begin{equation}
\ave{t_{ij}^+} = \left\{ \begin{array}{ l l}
\frac{ \ave{t_{ij}}}{1-e^{-\ave{t_{ij}}}} & \ave{t_{ij}} > 0 \\
0 & \ave{t_{ij}} = 0
\end{array} \right .
\end{equation}
which converges to the average value for highly used trips. Hence, the coefficient of determination $R^2_{cond}$ assuming an identity relation $ \ave{t_{ij}^+} \simeq t^{data}$ is explicitly,
\begin{equation}
R^2_{cond} = 1 - \frac{\sum_{ij|ij \in \mathcal{E}} \left(\ave{t_{ij}^{+,model}} - t_{ij}^{data}\right )^2}{\sum_{ij|ij \in \mathcal{E}} \left(\ave{t_{ij}^{+,model}} - \overline{\ave{t_{ij}^{+,model}}}\right)^2}.
\end{equation}

						\end{itemize}
	
		\paragraph{Network measures}
			To better grasp the quality of the models, we propose to compare also some of its multi-edge network related quantities:
			\begin{itemize}
				\item \textbf{Degree:} The unweighted degree of the nodes is the sum of their incoming/outgoing active edges (edges for which $t_{ij}>0$), $k^{x} = \sum_{x} \Theta(t_{ij})$ being $E = \sum_{ij} \Theta(t_{ij})$ the total number of active edges and $x$ referring to the outgoing $i$ or incoming direction $j$.
				\item \textbf{Average weighted neighbor strength:} This metric is widely used in the literature. It indicates the level of correlations at the node level and is defined as $s^w_{nn} (s^x_i) = \frac{\sum_{x} t_{ij} s^y_x}{s_i^x} $, where $y$ is the complementary of $x$ (if $x=i$ then $y=j$ and vice versa).
				\item \textbf{Distribution of weights on existing edges:} This commonly used measure is computed as $P(t) = \sum_{ij} \delta_{t,t_{ij}} / E$ and indicates the collection of weight values present in the network, where $\delta_{x,y}$ corresponds to the Kroenecker delta.
				\item \textbf{Graph average existing weight of trips as a function of product of incoming and outgoing degree}: To quantify the deviation from a completely randomized configuration model, we compute also this metric, which is the average weight of existing trips as a function of the product of out(in) strengths of their origin (destination) nodes: $\bar{t}(ss') = \sum_{ij | j = s',i=s} \hat{t}_{ij} / n_{ss'}$, where $n_{ss'}$ is the cardinality of the sum. For the configuration case, this magnitude is equivalent to $\ave{t^+} \propto \frac{s s'}{1- e^{-ss'/T}}$.
			\end{itemize}
		\paragraph{Information values}
We also assess the quality of our models using their Log-Likelihood values assuming a set of independent Poisson random variables with known means $\ave{t_{ij}}$ for each intersection pair,
			\begin{equation}
			 	\mathcal{L} = \ln P(\col{\hat{t_{ij}}} | \col{\ave{t_{ij}}}) = \sum_{ij} \ln \left( e^{-\ave{t_{ij}}} \frac{\ave{t_{ij}}^{\hat{t}_{ij}}}{\hat{t}_{ij}!} \right).
			\end{equation}
Incompatible Loglikelihood values are not reported in tables (such as cases where $\ave{t_{ij}} = 0 < \hat{t}_{ij}$ or $| \hat{t}_{ij} - \ave{t_{ij}}| \gg 0$).

	\subsection*{Simulations}
All the simulations and solving of saddle point equations as well as the analysis of the multi-edge networks have been performed using the freely available, open source package Origin-Destination Multi-Edge analysis (ODME)\cite{ODME2014}.


%

\section*{Acknowledgements}
We thank P. Colomer-de-Simon for useful comments and suggestions.


\begin{thebibliography}{10}

\bibitem{gonzalez2008uih}
Gonz{\'a}lez MC, Hidalgo CA, Barab{\'a}si AL.
\newblock Understanding individual human mobility patterns.
\newblock Nature. 2008;453(7196):779--782.

\bibitem{blondel2015}
Blondel VD, Decuyper A, Krings G.
\newblock A survey of results on mobile phone datasets analysis.
\newblock arXiv preprint arXiv:150203406. 2015;.

\bibitem{bazzani2010slu}
Bazzani A, Giorgini B, Rambaldi S, Gallotti R, Giovannini L.
\newblock Statistical laws in urban mobility from microscopic GPS data in the
  area of Florence.
\newblock J Stat Mech. 2010;2010:P05001.

\bibitem{brockmann2006slh}
Brockmann D, Hufnagel L, Geisel T.
\newblock The scaling laws of human travel.
\newblock Nature. 2006;439(7075):462--465.

\bibitem{thiemann2010sbs}
Thiemann C, Theis F, Grady D, Brune R, Dirk~Brockmann D.
\newblock The Structure of Borders in a Small World.
\newblock PLoS one. 2010;5:e15422.

\bibitem{scellato2011ssp}
Scellato S, Noulas A, Lambiotte R, Mascolo C.
\newblock Socio-spatial Properties of Online Location-based Social Networks.
\newblock Proceedings of ICWSM. 2011;11.

\bibitem{scellato2011nps}
Scellato S, Musolesi M, Mascolo C, Latora V, Campbell A.
\newblock NextPlace: A Spatio-Temporal Prediction Framework for Pervasive
  Systems.
\newblock Pervasive Computing. 2011;p. 152--169.

\bibitem{barthelemy2010sn}
Barth{\'e}lemy M.
\newblock Spatial Networks.
\newblock Phys Rep. 2010;499:1--101.

\bibitem{cattuto2010dpi}
Cattuto C, Van~den Broeck W, Barrat A, Colizza V, Pinton JF, Vespignani A.
\newblock Dynamics of person-to-person interactions from distributed RFID
  sensor networks.
\newblock PloS one. 2010;5(7):e11596.

\bibitem{roth2010cpc}
Roth C, Kang SM, Batty M, Barth{\'e}lemy M.
\newblock Structure of Urban Movements: Polycentric Activity and Entangled
  Hierarchical Flows.
\newblock PLoS ONE. 2011 01;6(1):e15923.

\bibitem{szell2012ums}
Szell M, Sinatra R, Petri G, Thurner S, Latora V.
\newblock Understanding mobility in a social petri dish.
\newblock Scientific Reports. 2012;2:457.

\bibitem{song2010msp}
Song C, Koren T, Wang P, Barab\'asi AL.
\newblock Modelling the scaling properties of human mobility.
\newblock Nature Physics. 2010 10;6:818--823.

\bibitem{song2010lph}
Song C, Qu Z, Blumm N, Barab\'asi AL.
\newblock Limits of predictability in human mobility.
\newblock Science. 2010;327(5968):1018.

\bibitem{hufnagel2004fce}
Hufnagel L, Brockmann D, Geisel T.
\newblock Forecast and control of epidemics in a globalized worlds.
\newblock Proc Natl Acad Sci USA. 2004;101:15124--15129.

\bibitem{belik2011nhm}
Belik V, Geisel T, Brockmann D.
\newblock Natural human mobility patterns and spatial spread of infectious
  diseases.
\newblock Phys Rev X. 2011;1:011001.

\bibitem{colizza2006rat}
Colizza V, Barrat A, Barth{\'e}lemy M, Vespignani A.
\newblock The role of the airline transportation network in the prediction and
  predictability of global epidemics.
\newblock Proc Natl Acad Sci USA. 2006;103(7):2015.

\bibitem{balcan2009mmn}
Balcan D, Colizza V, Goncalves B, Hu H, Ramasco JJ, Vespignani A.
\newblock Multiscale mobility networks and the spatial spreading of infectious
  diseases.
\newblock Proc Natl Acad Sci USA. 2009;106(51):21484--21489.

\bibitem{batty2013nsc}
Batty M.
\newblock The New Science of Cities.
\newblock MIT Press; 2013.

\bibitem{zipf1946p}
Zipf GK.
\newblock The P 1 P 2/D hypothesis: on the intercity movement of persons.
\newblock American sociological review. 1946;11(6):677--686.

\bibitem{simini2012umm}
Simini F, Gonz{\'a}lez MC, Maritan A, Barab{\'a}si AL.
\newblock A universal model for mobility and migration patterns.
\newblock Nature. 2012;484(7392):96--100.

\bibitem{stouffer1940iot}
Stouffer SA.
\newblock {Intervening Opportunities: A Theory Relating Mobility and Distance}.
\newblock American Sociological Review. 1940;5(6):845--867.

\bibitem{Clauset2009}
Clauset A, Shalizi CR, Newman MEJ.
\newblock Power-Law Distributions in Empirical Data.
\newblock SIAM Rev. 2009;51(4):661--703.

\bibitem{sagarra2011sca}
Sagarra O.
\newblock Statistical Complex Analysis of Taxi Mobility in San Francisco.
\newblock Universitat Polit{\`e}cnica de Catalunya; 2011.

\bibitem{Santi14}
Santi P, Resta G, Szell M, Sobolevsky S, Strogatz SH, Ratti C.
\newblock Quantifying the Benefits of Vehicle Pooling with Shareability
  Networks.
\newblock Proc National Academy of Science. 2014;111(37):13290--13294.

\bibitem{Piorkowski09}
Piorkowski M, Sarafijanovic-Djukic N, Grossglauser M.
\newblock A Parsimonious Model of Mobile Partitioned Networks.
\newblock In: IEEE Conference on Communication Systems and Networks (COMSNET).
  IEEE; 2009. p. 1--10.

\bibitem{Spieser14}
Spieser K, Treleaven K, Zhang R, Frazzoli E, Morton D, Pavone M.
\newblock Towards a Systematic Approach to the Design and Evaluation of
  Automated Mobility-on-Demand Systems: a Case Study in Singapore.
\newblock Road Vehicle Automation. 2014;p. 229--245.

\bibitem{wilson1970stp}
Wilson A.
\newblock {A statistical theory of spatial distribution models}.
\newblock demand Travel theory Exp. 1970;1(3):253--269.

\bibitem{erlander1990gmt}
Erlander S, Stewart NF.
\newblock The gravity model in transportation analysis : theory and extensions.
\newblock VSP Utrecht; 1990.

\bibitem{sagarra2013smm}
Sagarra O, P{\'e}rez-Vicente C, D{\'\i}az~Guilera A.
\newblock Statistical mechanics of multi-edge networks.
\newblock Physical Review E, 2013, vol 88, p 062806-1-062806-14. 2013;.

\bibitem{sagarra2014efn}
Sagarra O, Font-Clos F, P{\'e}rez-Vicente CJ, D{\'\i}az-Guilera A.
\newblock {The configuration multi-edge model: Assessing the effect of fixing
  node strengths on weighted network magnitudes}.
\newblock Europhysics Letters. 2014 Aug;107(3):38002.

\bibitem{Peng2012chm}
Peng C, Jin X, Wong K, Shi M, Lio P.
\newblock Collective Human Mobility Pattern from Taxi Trips in Urban Area.
\newblock PLoS ONE. 2012 04;7(4):e34487.

\bibitem{yang2014lpc}
Yang Y, Herrera C, Eagle N, Gonz{\'a}lez MC.
\newblock Limits of Predictability in Commuting Flows in the Absence of Data
  for Calibration.
\newblock Sci Rep. 2014 07;4.

\bibitem{bianconi2009a}
Bianconi G, Pin P, Marsili M.
\newblock Assessing the relevance of node features for network structure.
\newblock Proc Natl Acad Sci. 2009;106:11433--11438.

\bibitem{Lenormand2014}
Lenormand M, Picornell M, Cant\'{u}-Ros OG, Tugores A, Louail T, Herranz R,
  et~al.
\newblock {Cross-checking different sources of mobility information.}
\newblock PLoS One. 2014;9(8):e105184.

\bibitem{Lenormand2012}
Lenormand M, Huet S, Gargiulo F, Deffuant G.
\newblock {A universal model of commuting networks.}
\newblock PLoS One. 2012 Jan;7(10):e45985.

\bibitem{robillard1975eom}
Robillard P.
\newblock Estimating the OD matrix from observed link volumes.
\newblock Transportation Research. 1975;9(2):123--128.

\bibitem{ODME2014}
Sagarra O. ODME: Origin Destination Multi-Edge network package; 2014.
\newblock Available from: \url{https://github.com/osagarra/ODME_lite}.

\end{thebibliography}
%
%
%

\end{document}